\documentclass[numbib]{imaiai}

\usepackage{multirow}
\usepackage{verbatim}
\usepackage{color,graphicx}
\usepackage[numbers]{natbib}
\def\url#1{\expandafter\string\csname #1\endcsname}
\newcommand{\Proof}{\noindent\textbf{Proof.}\quad}

\newcommand{\norm}[1]{\left\lVert #1 \right\rVert}

\begin{document}

\title{Fast link prediction for large networks using spectral embedding}

\shorttitle{Fast link prediction with spectral embedding} %%%for recto running head
\shortauthorlist{Benjamin Pachev and Benjamin Webb} %%% for verso running head

\author{{%%%% First author details
\sc Benjamin Pachev, Benjamin Webb}$^*$\\[2pt]
\textit{Department of Mathematics, Brigham Young University, Provo, Utah}\\
$^*${\email{Corresponding author. Email: bwebb@mathematics.byu.edu}}
}

\date{\today}

%\pacs{03.67.Pp, 03.67.Lx}

\maketitle

\begin{abstract}
{Many link prediction algorithms require the computation of a similarity metric on each vertex pair, which is quadratic in the number of vertices and infeasible for large networks. We develop a class of link prediction algorithms based on a spectral embedding and the $k$ closest pairs algorithm that are scalable to very large networks. We compare the prediction accuracy and runtime of these methods to existing algorithms on several large link prediction tasks. Our methods achieve comparable accuracy to standard algorithms but are significantly faster.}
{link prediction; graph embedding; commute time; resistance distance; closest pairs.}  
\end{abstract}

\section{Introduction}
The study of networks has become increasingly relevant in our understanding of the technological, natural, and social sciences. This is owing to the fact that many important systems in these areas can be described in terms of networks \cite{networks_an_introduction}, where vertices represent the system's individual components, e.g. computer routers, neurons, individuals, etc. and where edges represent interactions or relationships between these components.

An essential feature of the large majority of these networks is that they have a dynamic topology, i.e. a structure of interactions that evolves over time \cite{adaptive_networks}. The structure of social networks, for instance, change over time as relationships are formed and dissolved. In information networks such as the WWW the network's structure changes as information is created, updated, and linked.

Although understanding the mechanisms that govern this structural evolution is fundamental to network science, these mechanisms are still poorly understood. Consequently, predicting a network's eventual structure, function, or whether the network is likely to fail at some point are all currently out of reach for even simple networks.

In an attempt to determine which processes cause changes in a network's structure we are lead to the following link prediction problem: Given a network, which of the \textit{links}, i.e. edges between existing vertices, are likely to form in the near future. Here we adopt the standard convention that links are to be predicted solely on the basis of the network's current topology (see, for instance, \cite{link_prediction_social_networks}).

Importantly, the link prediction problem can be used to study more than just which edges will appear in a network. It can also be used to predict which of the non-network edges are, in fact, in the network but currently undetected. Similarly, it can be used to detect which of the current network edges have been falsely determined to be a part of the network.

This notion of link prediction is of central importance in numerous applications. Companies such as Facebook, Twitter, and Google need to know the current state and efficiently predict the future structure of the networks they use to accurately sort and organize data \cite{quercia2012tweetlda}. Biologists need to know whether biochemical reactions are caused by specific sets of enzymes to infer causality and so on \cite{barzel2013network}.

The barrier in determining whether network links truly exist in these and other settings, is that testing and discovering interactions in a network requires significant experimental effort in the laboratory or in the field \cite{clauset2008hierarchical}. Similarly, determining experimentally when and where a new link will form may also be impractical, especially if the precise mechanism for link formation is unknown. For these reasons it is important to develop models for link prediction.

At present, there is an ever increasing number of proposed methods for predicting network links \cite{srinivas2016link}. Not surprisingly, certain methods more accurately predict the formation of links in certain networks when compared with others. Additionally, each of these methods has a runtime that scales differently with the size of the network. In our experiments, we discover that a number of link predictors have a runtime that is so high that it effectively prohibits their use on moderately large networks.

Here we propose a class of link predicting algorithms that scale to large networks. This method, which we refer to as the \textit{approximate resistance distance predictor}, integrates a spectral embedding of the network with a known algorithm for efficiently finding the $k$ closest pairs of points in Euclidean space. The spectral embedding aspect of the algorithm is derived as a low-rank approximation of the effective resistance between network vertices, as in \cite{fouss2007random}. The $k$ closest pairs component of the algorithm is taken from \cite{lenhof1992k} and can be used to predict links based on this embedding.

Here we compare the prediction accuracy and runtime of this method against several well-known algorithms on a number of coauthorship networks and a social network consisting of a small subset of Facebook users. We find that our method is achieves the best accuracy on some networks and scales to networks that many other link predictors cannot.

The paper is structured as follows. In Section \ref{sec:linkpred} we describe the link prediction problem and outline a number of standard link prediction algorithms. In Section \ref{sec:spectral} we introduce the method of resistance distance embedding and prove that it is optimal as a low rank approximation of effective resistance (see Proposition \ref{prop:resistance_distance_approx}). In Section \ref{sec:setup} we describe the experimental setup. Section \ref{sec:results} numerical results comparing the performance of the resistance distance embedding algorithm to other algorithms are given. Section VI concludes with some closing remarks including a number of open questions for future work.

\section{The Link Prediction Problem}\label{sec:linkpred}
 The link prediction problem can be stated as follows. Given a connected graph $G = (V, E)$, and $k$, the number of predicted nonadjacent links, we seek $k$ pairs of vertices which are most likely to become connected. While the choice of $k$ depends on the application, we adopt the convention that $1 \leq k \leq |E|$.

The general paradigm for link prediction is to compute a similarity metric $score(x,y)$ on each vertex pair $(x,y)$. The predicted links are then the $k$ $(x,y) \in V \times V-E$ for which $score(x,y)$ is maximal. By contructing a matrix from the scores, we obtain a \textit{graph kernel}. We can also go in the other direction. Any real $n \times n$ matrix, where $n = |V|$, defines a score function on pairs of vertices, and can be used for link prediction.

We now give a sampling of existing link prediction algorithms.

\subsection{Local Methods}
A \textit{local method} for link prediction is an algorithm that uses vertex neighborhoods to compute similarity. 

\textbf{Common Neighbors:}
Common neighbors simply assigns 
\begin{equation}\label{eq:common}score(x,y) = |\Gamma(x)\cap\Gamma(y)|,
\end{equation}
 where $\Gamma(x)$ is the neighbor set for $x \in V$.

\textbf{Jaccard's Coefficient:}
Jaccard's coefficient is a normalized version of common neighbors that takes into account the total number of neighbors for both vertices. It is given by 
\begin{equation}\label{eq:jaccard}score(x,y) = \frac{|\Gamma(x)\cap\Gamma(y)|}{|\Gamma(x)\cup\Gamma(y)|}.
\end{equation}

\textbf{Preferential Attachment:}
Preferential attachement is based on the idea that highly connected nodes are more likely to form links, an observed pattern in coathourship networks \cite{newman2001clustering}. This leads to 
\begin{equation}\label{eq:prefattach}score(x,y) = |\Gamma(x)||\Gamma(y)|.
\end{equation}

\textbf{Adamic-Adar:}
\begin{equation}\label{eq:adamic}score(x,y) = \sum_{z \in \Gamma(x)\cap\Gamma(y)}\frac{1}{log|\Gamma(z)|}
\end{equation}

\textbf{Resource Allocation:}
\begin{equation}\label{eq:resource_alloc}score(x,y) = \sum_{z \in \Gamma(x)\cap\Gamma(y)}\frac{1}{|\Gamma(z)|}
\end{equation}

\subsection{Path-based Methods}
\textit{Path-based methods} consider all or a subset of the paths between two vertices to compute similarity. Unlike local similarity measures, they can capture global information about the network.

\textbf{Shortest Path:}
This link predictor defines $score(x,y)$ as the negated length of the shortest path from $x$ to $y$.

\textbf{Katz:}
The Katz metric counts all paths between two nodes, and discounts the longer paths exponentially. Define $path^{\ell}_{x,y}$ to be the set of all paths of length $\ell$ from $x$ to $y$. Then given a weight $0 < \beta < 1$,

\begin{equation}\label{eq:katz}score(x,y) = \sum_{\ell=1}^{\infty}\beta^{\ell}|path^{\ell}_{x,y}|
\end{equation}

A closed form for the associated graph kernel is given by $(I-\beta A)^{-1} - I$ = $\sum_{\ell=1}^{\infty}(\beta A)^{\ell}$, where $A$ is the adjacency matrix of $G$.
	
\subsection{Random walks}
A \textit{random walk} on G starts at some node x and iteratively moves to new nodes with uniform probability. There are a multitude of link predictors based on random walks. These are some of the fundamental ones.

\textbf{Hitting and Commute Time:} The \textit{hitting time} $H_{x,y}$ is the expected number of steps required to reach $y$ in a random walk starting at $x$. Commute time is defined as $C_{x,y} = H_{x,y} + H_{y,x}$. Negated hitting time can be used as a link predictor, but the hitting time is assymetric in general, so we use instead the negated commute time, which is symmetric.

The commute time and its variants will be discussed further in Section \ref{sec:spectral}.

\textbf{Rooted Page Rank:}
A problem with hitting and commute time is that random walks can become lost exploring distant portions of the graph. Rooted Page Rank deals with this problem by introducing random resets. Given a root node $x$, we consider a random walk starting at $x$. At each step, with probability $\alpha$ the walk returns back to $x$. With probability $1-\alpha$ the walk proceeds to a random neighbor. Given a root node $x$, for each other node $y$, $score(x,y)$ is defined as the stationary probability of $y$ under the random walk rooted at $x$. The corresponding graph kernel is given by $(1-\alpha)(I-\alpha D^{-1}A)^{-1}$, where $D$ is the degree matrix and $A$ is the adjacency matrix.

\subsection{Scaling Link Predictors to Large Networks}

 Many link predictors, such as Katz, require the computation of a matrix inverse. This is heinously expensive for large networks, as it is cubic in the number of vertices. One way to circumvent such problems is via a low-rank approximation of the score matrix. We investigate such a low-rank approximation for the commute-time or resistance distance kernel in the next section.  

 Even the simpler local predictors such as common neighbors or preferential attachment face difficulties at scale. This is because for sufficiently large networks, it is not possible to compute scores for each pair of vertices and then find the maximal ones. Instead, efficient search techniques must be employed to search only a small subset of the potential links in order to find those of maximal score. In Section \ref{sec:spectral} we will demonstrate how a class of graph embedding based predictors can efficiently find the $k$ links of maximal score.

\section{Spectral Embedding}\label{sec:spectral}

We begin by deriving the \textit{approximate resistance distance link predictor} as a best low-rank approximation to commute time and show how to evaluate its link prediction scores with a spectral embedding. We then show that this link predictor is part of a family of graph embedding based link predictors that use the $k$ closest pairs algorithm to efficiently find the links of maximal score. Finally, we discuss efficient ways to compute the spectral embedding upon which the approximate resistance distance predictor relies.

\subsection{Approximating Commute Time}
Let $L$ = $D-A$ be the Laplacian matrix of a graph $G=(V,E)$, and let $n$ = $|V|$.  Let $L^{\dagger}$ be the Moore-Penrose inverse of L. Then the commute time is given by
\begin{equation}\label{eq:commute_time}C_{x,y} = |E|(L^{\dagger}_{x,x} + L^{\dagger}_{y,y} - 2L^{\dagger}_{x,y}),
\end{equation} where the quantity $r_{x,y} = (L^{\dagger}_{x,x} + L^{\dagger}_{y,y} - 2L^{\dagger}_{x,y})$ is known as the \textit{effective resistance} or the \textit{resistance distance} \cite{fouss2007random}. Since resistance distance differs from commute-time by a (network-dependant) constant scaling factor, they can be used interchangeably for link prediction.

For many networks, $G$ is too large to compute $L^{\dagger}$ exactly, so an approximation must be used. A natural choice is a best rank-$d$ approximation to $L^{\dagger}$ for some fixed dimension $d$. The resulting approximation of the resistance distances is closely related to distances between points in Euclidean space.

\begin{proposition}\label{prop:resistance_distance_approx}Let $d$ be a positive integer and let $G=(V,E)$ be a connected, undirected graph. Then $\exists$ a best rank-$d$ approximation $S$ of $L^{\dagger}$, and a map $f:V\rightarrow R^{d}$ so that $\forall $ $x,y \in V$, $S_{x,x} + S_{y,y} - 2S_{x,y} = \norm{f(x)-f(y)}_2^2$. We call this map the \textit{resistance distance embedding}. \end{proposition}
$\Proof$
 For a connected graph, the Laplacian matrix is positive semidefinite, with eigenvalues $0=\lambda_1 < \lambda_2 \leq \dots \leq \lambda_n$ and corresponding eigenvectors $v_{1}, v_{2}, v_{3}, \dots, v_{n}$. Then we have the spectral decompositions $$L = \sum_{i=2}^{n}\lambda_{i}v_{i}v_{i}^T$$
and
$$L^{\dagger} = \sum_{i=2}^{n}\frac{1}{\lambda_{i}}v_{i}v_{i}^T.$$
Hence, $S = \sum_{i=2}^{d+1}\frac{1}{\lambda_{i}}v_{i}v_{i}^T$ is a best rank-$d$ approximation to $L^{\dagger}$ in the 2-norm.
 Then note $$S_{x,x} + S_{y,y} - 2S_{x,y} = (e_{x}-e_y)^{T}S(e_{x}-e_y) $$
 $$ = \sum_{i=2}^{d+1}\frac{1}{\lambda_{i}}(e_{x}-e_y)^{T}v_{i}v_{i}^T(e_{x}-e_y) $$
 $$ = \sum_{i=2}^{d+1}\frac{1}{\lambda_{i}}(v_{i,x}-v_{i,y})^2$$
 $$ = \norm{f(x)-f(y)}_2^2$$
 where 
 \begin{equation}\label{eq:res_embedding}f(x) = [\frac{v_{2,x}}{\sqrt{\lambda_2}}, \frac{v_{3,x}}{\sqrt{\lambda_3}}, \dots , \frac{v_{d+1,x}}{\sqrt{\lambda_{d+1}}}]^{T} \in R^{d}
 \end{equation}
\qed

We define the \textit{approximate resistance distance link predictor} of dimension $d$ by setting 
\begin{equation}\label{eq:approx_predictor}score(x,y) = -(S_{x,x} + S_{y,y} - 2S_{x,y}) = -\norm{f(x)-f(y)}_2^2,
\end{equation} where $S$ and $f$ are defined as in Proposition \ref{prop:resistance_distance_approx}.

In the next section, we will see that the approximate resistance distance link predictor is part of a class of link predictors that avoid brute-force search when predicting links.

\subsection{Link Prediction with Graph Embeddings}
The resistance distance embedding is a special case of a \textit{graph embedding}, which is a map $f$ from $V$ to $R^d$,  $d$ a positive integer. We can use graph embeddings to create link predictors. A natural choice is to set $score(x,y)$ =  $-\norm{f(x)-f(y)}_2$, (so maximizing score corresponds to minimizing distance). We refer to this score function as the \textit{Euclidean score}.

 If $f$ is the resistance distance embedding, then link prediction with the Euclidean score is equivalent to the approximate resistance distance predictor. Recall that the approximate resistance distance score function is $-\norm{f(x)-f(y)}_2^2$. The $k$ predicted links of maximal score correspond to the $k$ nonadjacent pairs of vertices $(x,y)$ for which $-\norm{f(x)-f(y)}_2^2$ is maximal. These are precisely the $k$ links for which $\norm{f(x)-f(y)}_2$ is minimal and are predicted with the Euclidean score.

 Link prediction with the Euclidean score is related to the $k$ \textit{closest pairs problem}. The closest pairs problem is as follows. Given a set of vectors $\{x_1,x_2, \dots, x_n\} \subset R^d$ we seek the k unordered pairs $(x_i, x_j), i \ne j$ of minimal distance (here we use the Euclidean norm but any $L_{p}$ norm can be used, $1 \leq p \leq \infty$). There is an algorithm to solve this problem in 
 \begin{equation}\label{eq:k_closest}O(d(n\,log\,n + k\,log\,n\,\,log(\frac{n^2}{k})))
 \end{equation} \citep{lenhof1992k}.

  We can think of the link prediction problem as the closest pairs problem applied to the set of vectors $\{f(y), y \in V\}$, with the additional constraint that the best pairs must correspond to non-edges in $G$. The extra constraint can be handled by finding the $|E|+k$ closest pairs, then selecting the best $k$ which are non-edges. As there can be no more than $|E|$ edges, this approach is sure to work. We then have the worst-case complexity bound of
  \begin{equation}\label{eq:link_pred_complexity}O(d(n\,log\,n + (|E|+k)log\,n\,\,log(\frac{n^2}{|E|+k}))).
  \end{equation} 

Recalling that we require $1 \leq k \leq |E|$, and assuming that $G$ is connected so $|E| \geq n-1$, this complexity bound can be simplified to 
   \begin{equation}\label{eq:link_pred_simple_complexity}O(d\,|E|\,log^2n).
   \end{equation}
  For large, sparse networks, $|E| << n^2$, and this is a tremendous speedup over the $O(n^2)$ brute-force approach. 

\textbf{Cosine Similarity Score:}\,
Another link prediction score function that can be derived from a graph embedding is the cosine similarity score, defined by \begin{equation}\label{eq:cosine_similarity}score(x,y)=\frac{<f(x),f(y)>}{\norm{f(x)}\norm{f(y)}}.
\end{equation}

If the cosine similarity score is used, the link prediction problem can still be solved without brute-force search. It is equivalent to the link prediction problem with Euclidean score on a modified graph embedding. The modified embedding is obtained from the original by normalizing the embedding vectors as follows.
 \begin{proposition}\label{prop:cosine_score}
Given a graph embedding $f:V\rightarrow R^d$, the link prediction problem using $$score(x,y)=\frac{<f(x),(f(y))>}{\norm{f(x)}\norm{f(y)}} = \cos\theta$$ is equivalent to the link prediction problem with the Euclidean score function on the modified embedding given by $g(y) = \frac{f(y)}{\norm{f(y)}}$.
\end{proposition}
$\Proof$
Let $x,y \in V$. Note $$<g(x), g(y)> = \cos\theta = score(x,y).$$
We have
$$\norm{g(x)-g(y)}_{2}^2 = \norm{g(x)}_{2}^2 + \norm{g(y)}_{2}^2 - 2<g(x), g(y)>$$
$$= 2 - 2\cos\theta = 2-2score(x,y).$$ This shows that minimizing Euclidean distance for the modified embedding is the same as maximizing cosine similarity score on the original, so link prediction with Euclidean score on the modified embedding is equivalent to link prediction with the cosine similarity score on the original.
\qed

This section introduced a class of link predictors that avoid a brute-force search when predicting links. These link predictors rely on a precomputed graph embedding. The graph embedding needs to be efficiently computable in order for the overall prediction algorithm to be fast.
We are concerned with link predictors that rely on the resistance distance embedding. Consequently, rapid computation of this particular graph embedding is the subject of the next section.

\subsection{Computing the Resistance Distance Embedding}

Computing the resistance distance embedding of dimension $d$ requires finding the smallest $d$ nonzero eigenvalues and associated eigenvectors of the Laplacian matrix $L$. Fortunately, specialized, efficient algorithms exist for this problem which exploit the positive semi-definiteness and sparsity of $L$. These include TRACEMIN-Fiedler \cite{manguoglu2010tracemin} and a multilevel solver MC73\_FIEDLER \cite{hu2003hsl}. TRACEMIN-Fiedler is simpler to implement, and is also parallelizable, so we use it in our experiments.

\section{Experimental Setup}\label{sec:setup}

In this section we compare the performance of our link prediction algorithm to others on several large social networks. In a social network, nodes correspond to persons or entities. Edges correspond to an interaction between nodes, such as coauthouring a paper or becoming linked on a social media website.

\subsection{The Networks}

\textbf{Arxiv High Energy Physics Theory (hep-th):}
This network is a coauthorship network obtained from the Konect network collection. \cite{konect:leskovec107, konect:2016:ca-cit-HepTh}.

\textbf{Arxiv High Enery Physics Phenomenology (hep-ph):}
This is another coauthorship network from the Konect network collection \cite{konect:leskovec107, konect:2016:ca-cit-HepPh}.

\textbf{Facebook Friendship (facebook):}
This social network consists of a small subset of facebook users, where edges represent friendships \cite{konect:2016:facebook-wosn-links, viswanath09}. 

\textbf{Arxiv Condensed Matter Physics (cond-mat):}
This dataset was obtained from Mark Newman's website \cite{newman2001structure}, and is also a coathourship network. Unlike the other datasets, the edges are not timestamped.  

\subsection{Creating Training Graphs}
    In order to perform link prediction, we partition edges into a training set and a test set. Edges in the training set occur before those in the test set and are used to construct a training graph. We run link prediction algorithms on the training graph to predict the contents of the test set. In most cases, edges have timestamps, and we can choose a cutoff time to partition the edges. 

	For one network (cond-mat) the edges are not timestamped. However, there are two versions of the cond-mat network available. One contains all collaborations up to 2003. The second is an updated network with all collaborataions up to 2005. We use the first network as the training graph. The test set consists of all edges in the second network for which both nodes are in the earlier network. 
 
    Choosing the cutoff between the training and test edges is somewhat arbitrary. If too few edges are used for training, link predictors will struggle. If too few are left for testing, then results may be statistically insignificant. See Table \ref{table:train_net_stats} for a comparison of the training networks and original networks.
\begin{table}[h!]
\centering
\begin{tabular}{||c c c c||} 
 \hline
 Network & Nodes & Edges & Average Degree \\ [0.5ex] 
 \hline\hline
 cond-mat & 15,803 & 60,989 & 7.7187 \\
 cond-mat train & 13,861 & 44,619 & 6.4381 \\
 facebook & 63,731 & 817,035 & 12.8201 \\ 
 facebook train & 59,416 & 731,929 & 24.6374 \\
 hep-ph & 28,093 & 3,148,447 & 112.0723 \\
 hep-ph train & 26,738 & 2,114,734 & 158.1819 \\
 hep-th & 22,908 & 2,444,798 & 106.7225 \\
 hep-th train & 21,178 & 1,787,157 & 168.7749 \\ [1ex] 
 \hline
\end{tabular}
\caption{Training network statistics}
\label{table:train_net_stats}
\end{table}
    	
Our spectral embedding based link prediction algorithms require a connected graph. To solve this problem, we reduce each training graph to its largest connected component. For each network we consider, the largest component contains the vast majority of the vertices.

\subsection{The Predictors}

We perform experiments with two spectral embedding based predictors. Each uses the resistance distance embedding of dimension $d$, with $d$ a parameter to be varied. The first uses the Euclidean score function and is equivalent to the approximate resistance distance predictor of dimension $d$. The second uses the cosine similarity score. We refer to these link predictors as spec\_euclid and spec\_cosine respectively (spec for spectral). In tables, the dimension of the embedding is indicated by a number after the predictor name. For example, spec\_euclid8 refers to the spec\_euclid predictor using an 8-dimensional resistance distance embedding. 

The other link prediction algorithms used in our experiments are preferential attachment, common neighbors, Adamic Adar, Rooted Page Rank and Katz (with $\beta=.01$). Some networks are too large for certain algorithms to handle, so not every algorithm is run on each network. For example, the facebook training graph has 59,416 nodes. Computing the Katz score on this graph requires finding the inverse of a 59,416$\times$59,416 matrix, and is very expensive in time and space, so we do not use the Katz algorithm for the facebook graph.

All experiments were performed on the same 4 core machine. The common neighbors, preferential attachment, and Adamic Adar algorithms were implemented in Python and were not parallelized. Our spectral link predictors, Katz, and Rooted Page Rank use the Python library Numpy to parallelize linear algebra operations. All code that was used in the experments in this paper can be found at the git repository bitbucket.org/thorfax/spectral\_research.

	For each network, we fix the number of links to be predicted. With the exception of hep-th, this number is equal to 10\% of the maximum possible number of correct predictions (i.e the number of new links in the test set). For the hep-th network we discovered that the spec\_euclid and spec\_cosine predictors achieve nearly perfect accuracy when predicting 1000 links. As this is not the case for any other network we considered, we report this unusual phenomenon.
	
	For all of the networks we consider, the probability of randomly predicting a correct link is very low. Most of the algorithms we consider do much better than the random baseline, but have low raw accuracy since there are few new links compared to the number of possible links. See Table \ref{table:train_prob_info} for a summary of the number of links predicted and baseline probability of randomly predicting a correct link.

\begin{table}[h!]
\centering
\begin{tabular}{||c c c c||} 
 \hline
 Network & Links  & Random Accuracy (\%) \\ [0.5ex] 
 \hline\hline
 cond-mat & 1190 & 0.012 \\
 facebook & 7858 & 0.004\\ 
 hep-ph & 101466 & 0.286 \\
 reduced hep-ph & 1988 & 0.661 \\
 hep-th & 1000 & 0.296 \\
 reduced hep-th & 135 & 0.084 \\ [1ex] 
 \hline
\end{tabular}
\caption{Link Prediction Task Setup}
\label{table:train_prob_info}
\end{table}

\section{Results}\label{sec:results}

On the cond-mat and facebook networks, both the spec\_euclid and spec\_cosine predictors performed worse than the simple common neighbors predictor. In addition to the full networks, we also compared predictor accuracy on reduced versions of the hep-th and hep-ph networks, because the full networks are too large for methods like Katz, common neighbors, and Rooted Page Rank to complete in a reasonable amount of time. On our reduced version of the hep-th network, our embedding-based predictors did better than common neighbors but not as well as the Rooted Page Rank predictor. On the reduced hep-ph network, the spec\_euclid predictor performed significantly better than all other competitors, including our other embedding-based predictor, spec\_cosine. 

As Table \ref{table:cond-mat} shows, the best predictors for the cond-mat network were Katz and common neighbors.  Note that for both spec\_euclid and spec\_cosine, the accuracy increases with the dimension of the embedding.

\begin{table}[h!]\centering
\begin{tabular}{||c c c c||} 
 \hline
  Predictor & Correct (\%) & Time (s)\\ [0.5ex]
 \hline\hline
katz & 5.97 & 62.96 \\
commonNeighbors & 5.97 & 1.55 \\
prefattach & 1.93 & 0.35 \\
spec\_euclid1 & 1.51 & 2.99 \\
spec\_cosine1 & 0.25 & 3.35 \\
spec\_euclid2 & 1.51 & 3.65 \\
spec\_cosine2 & 1.18 & 3.80 \\
spec\_euclid4 & 1.76 & 10.54 \\
spec\_cosine4 & 1.34 & 10.89 \\
spec\_euclid8 & 1.68 & 11.41 \\
spec\_cosine8 & 1.34 & 10.73 \\
spec\_euclid16 & 1.68 & 29.91 \\
spec\_cosine16 & 1.43 & 32.31 \\
  [1ex] 
 \hline
\end{tabular}
\caption{Performance of link predictors on the cond-mat network}
\label{table:cond-mat}
\end{table}

As previously mentioned, the facebook graph was too large to run the Katz predictor on it in a reasonable amount of time. As with the cond-mat network, the simple common neighbors predictor performs best.

\begin{table}[h!]\centering
\begin{tabular}{||c c c c||} 
 \hline
  Predictor & Correct (\%) & Time (s)\\ [0.5ex]
 \hline\hline
commonNeighbors & 5.29 & 151.76 \\
prefattach & 0.41 & 7.00 \\
spec\_euclid1 & 0.42 & 9.86 \\
spec\_cosine1 & 0.00 & 47.20 \\
spec\_euclid2 & 0.50 & 11.52 \\
spec\_cosine2 & 0.42 & 12.96 \\
spec\_euclid4 & 1.40 & 24.05 \\
spec\_cosine4 & 1.02 & 25.96 \\
spec\_euclid8 & 1.95 & 61.21 \\
spec\_cosine8 & 2.58 & 62.27 \\
  [1ex]
 \hline
\end{tabular}
\caption{Performance of link predictors on the facebook network}
\label{table:facebook}
\end{table}

Our spectral embedding link predictors performed significantly better on the hep-th and hep-ph networks, as Table \ref{table:hep-th} and Table \ref{table:hep-ph} show.

The common neighbors algorithm did not scale to the hep-th and hep-ph networks, unlike the facebook network. Although the facebook network had more nodes, it has a lower average node degree and fewer distance two pairs. The common neighbors algorithm computes intersections of neighbor sets for each distance two pair. Because the average node degree is higher for the hep-th and hep-ph networks, these intersections are more expensive to compute and there are more distance two pairs for which intersections must be computed.

\begin{table}[h!]\centering
\begin{tabular}{||c c c c||} 
 \hline
  Predictor & Correct (\%) & Time (s)\\ [0.5ex]
 \hline\hline
prefattach & 0.00 & 9.47 \\
spec\_euclid1 & 94.50 & 10.88 \\
spec\_cosine1 & 1.50 & 17.17 \\
spec\_euclid2 & 98.70 & 17.88 \\
spec\_cosine2 & 99.60 & 15.97 \\
spec\_cosine4 & 100.00 & 15.67 \\
spec\_euclid4 & 99.90 & 18.53 \\
spec\_euclid8 & 100.00 & 29.81 \\
spec\_cosine8 & 100.00 & 29.55 \\
spec\_euclid16 & 99.90 & 96.47 \\
spec\_cosine16 & 99.90 & 100.93 \\
  [1ex] 
 \hline
\end{tabular}
\caption{Performance of link predictors on the hep-th network}
\label{table:hep-th}
\end{table}

\begin{table}[h!]\centering
\begin{tabular}{||c c c c||} 
 \hline
  Predictor & Correct (\%) & Time (s)\\ [0.5ex]
 \hline\hline
prefattach & 0.00 & 16.85 \\
spec\_euclid1 & 3.93 & 18.25 \\
spec\_cosine1 & 0.14 & 32.37 \\
spec\_euclid2 & 9.16 & 21.98 \\
spec\_cosine2 & 3.44 & 23.46 \\
spec\_euclid4 & 19.25 & 27.04 \\
spec\_cosine4 & 13.65 & 26.09 \\
spec\_euclid8 & 22.90 & 46.69 \\
spec\_cosine8 & 21.12 & 49.62 \\
spec\_euclid16 & 24.62 & 148.51 \\
spec\_cosine16 & 23.97 & 135.83 \\
  [1ex] 
 \hline
\end{tabular}
\caption{Performance of link predictors on the hep-ph network}
\label{table:hep-ph}
\end{table}

In order to compare the performance of our spectral predictors to other predictors on the hep-ph and hep-th network data, we conducted another experiment using downsampled versions of these networks. To downsample, we used only the top 10\% highest degree nodes. Our spectral predictors performed the best on the reduced hep-ph network (see Table \ref{table:hep-ph small}), while the Rooted Page Rank algorithm was best for the reduced hep-th network (see Table \ref{table:hep-th small}).

\begin{table}[h!]\centering
\begin{tabular}{||c c c c||} 
 \hline
  Predictor & Correct (\%) & Time (s)\\ [0.5ex]
 \hline\hline
prefattach & 0.00 & 1.52 \\
katz & 1.16 & 2.75 \\
commonNeighbors & 9.36 & 23.96 \\
pageRank & 11.87 & 2.94 \\
adamicAdar & 8.85 & 754.40 \\
spec\_euclid1 & 1.81 & 2.80 \\
spec\_cosine1 & 0.10 & 6.21 \\
spec\_euclid2 & 4.73 & 3.73 \\
spec\_cosine2 & 2.57 & 6.86 \\
spec\_euclid4 & 13.13 & 5.80 \\
spec\_cosine4 & 11.12 & 9.01 \\
spec\_euclid8 & 16.40 & 7.39 \\
spec\_cosine8 & 9.31 & 7.44 \\
spec\_euclid16 & 14.13 & 17.32 \\
spec\_cosine16 & 4.93 & 14.90 \\
  [1ex] 
 \hline
\end{tabular}
\caption{Performance of link predictors on the reduced hep-ph network}
\label{table:hep-ph small}
\end{table}

\begin{table}[h!]\centering
\begin{tabular}{||c c c c||} 
 \hline
  Predictor & Correct (\%) & Time (s)\\ [0.5ex]
 \hline\hline
prefattach & 0.00 & 1.28 \\
katz & 0.00 & 1.61 \\
commonNeighbors & 2.22 & 16.70 \\
pageRank & 11.11 & 1.97 \\
adamicAdar & 2.22 & 788.11 \\
spec\_euclid1 & 0.00 & 2.02 \\
spec\_cosine1 & 0.00 & 3.77 \\
spec\_euclid2 & 0.74 & 4.45 \\
spec\_cosine2 & 0.00 & 4.61 \\
spec\_euclid4 & 0.74 & 6.84 \\
spec\_cosine4 & 0.00 & 6.25 \\
spec\_euclid8 & 2.22 & 10.59 \\
spec\_cosine8 & 1.48 & 11.02 \\
spec\_euclid16 & 8.89 & 26.82 \\
spec\_cosine16 & 5.93 & 24.92 \\
  [1ex] 
 \hline
\end{tabular}
\caption{Performance of link predictors on the reduced hep-th network}
\label{table:hep-th small}
\end{table}

\section{Conclusion}\label{sec:conclusion}

We present a link prediction framework that can scale to very large networks by avoiding the quadratic costs inherent in methods that exhaustively search all candidate pairs of nonadjacent nodes. We investigated the performance of a set of predictors based on this framework and the spectrum and eigenvectors of the graph's Laplacian matrix. These methods achieved high levels of accuracy on certain real-world link prediction tasks, and scaled well to networks with tens of thousands of nodes and millions of edges.

We emphasize that there are many other possible graph embeddings to invesitigate. Virtually all the runtime of our spectral link predictors is spent computing the resistance distance embedding. The $k$ closest pairs component of our algorithm is very fast in practice, with nearly linear temporal complexity in the number of edges Replacing the resistance distance embedding with one that is cheaper to compute could potentially produce link predictors that can scale to much larger networks than the ones we consider in this paper.

Our approximate resistance distance link predictor was derived as a low-rank approximation of resistance distance, an established link prediction score that is expensive to compute. Many other well-known predictors are expensive to compute, such as Katz and Rooted Page Rank. There is much room to explore low-rank approximations of such predictors and investigate whether they can be converted into accurate, scalable, graph embedding based, link predictors of the form we considered. 
\\
\\
\textbf{Funding}\\
This work was supported by the Defense Threat Reduction Agency [grant number HDTRA1-15-1-0049].

%\bibliographystyle{imaiai}
%\bibliography{imaiai_link_predrefs}
\ifx\undefined\BySame
\newcommand{\BySame}{\leavevmode\rule[.5ex]{3em}{.5pt}\ }
\fi
\ifx\undefined\textsc
\newcommand{\textsc}[1]{{\sc #1}}
\newcommand{\emph}[1]{{\em #1\/}}
\let\tmpsmall\small
\renewcommand{\small}{\tmpsmall\sc}
\fi

\end{document}